# First Demonstration of Multi-Agent LLM System for Million-Scale Optical Link Management in Global Production AIDCs

Jingyi Su[(1)], Yihao Zhang[(1)], Dianxuan Fu[(1)], Leiyan Fei[(1)], Juan Wang[(2)], Mengfan Dai[(2)], Qing Liu[(2)], Xiong Wu[(2)], Yufeng Jiang[(2)], Cheng Chen[(2)], Bowen Zhang[(2)*], Peilong Wang[(2)], Xi Chen[(3)], Zonglong He[(3)], Hongchen Yu[(3)], Zhicheng Ye[(3)], Weisheng Hu[(1)], and Qunbi Zhuge[(1)*]

[(1)]State Key Laboratory of Photonics and Communications, School of Information Science and Electronic Engineering, Shanghai Jiao Tong University, Shanghai, 200240, China, *qunbi.zhuge@sjtu.edu.cn
[(2)] Systems Department, Baidu, Beijing, China, *zhangbowen@baidu.com
[(3)] Optical Research Department, Huawei Technologies, Dongguan, China.

**Abstract** *We present the first LLM-powered multi-agent system for autonomous fault management across millions of optical links in production AIDCs. Refined via SFT and continuous memory evolution, it achieves 97.7% F1 and over 60% fault-incident reduction, outperforming SOTA LLMs on a ten-week field data evaluation.* 

## Introduction

The rapid advancements in large language models (LLMs) have driven AI data centers (AIDCs) to become the core of global computing infrastructure, where the requirements for network stability and reliability far exceed those of traditional DCs [1]. As GPU clusters scale toward the million-GPU level, even millisecond- or microsecond-level optical signal degradations or transient link flaps in high-speed optical interconnects can lead to hours of inference latency or training interruptions [2], resulting in massive economic losses and inflated operating expenses. Consequently, establishing a robust failure management system has become critical to ensuring the stable and efficient operation for AIDC networks [3].

Recently, data-driven techniques have been leveraged to facilitate troubleshooting and automate certain workflows in data center networks (DCNs) [4]. However, these approaches still require engineers to manually inspect multiple metrics and logs when port alarms are triggered, thereby limiting both flexibility and efficiency. Concurrently, LLMs exhibit superior proficiency in handling complex tasks over large-scale data [5-6]. Integrating LLM agents to safeguard operational AI training infrastructure within production environments presents a promising avenue that remains unexplored in current literature.

In this work, we present the first production network demonstration of OptiMIND (Optical Multi-agent Intelligent Network Diagnoser), an LLM-driven framework for autonomous AIDC network failure management. OptiMIND orchestrates a collaborative multi-agent architecture founded on a domain-adapted LLM augmented with a persistent memory system. To optimize diagnostic reasoning, we establish a self-evolving mechanism utilizing seven months of operational data for supervised fine-tuning (SFT) and continuous memory enrichment through operationally grounded feedback. Field validations across Baidu's production AIDC networks, spanning ten weeks and millions of optical links, demonstrate a 97.7% F1 score in failure prediction, while reducing the number of fault events by over 60%, establishing a reliable safeguard for large-scale AI infrastructure.

## Link Failure Incident Processing

Fig. 1(a) illustrates the global network infras-tructure in our demonstration that supports a wide range of AI services, including distributed computing and LLM training/inference. As shown in Fig. 1(b), the optical interconnect platform comprises several million transceivers, covering 100G and 400G rates over both single-mode and multi-mode fibers. Per-link performance metrics from both local-end and remote-end transceivers are collected over a 14-day observation window, including DDM, pre-FEC BER, SNR, and FDR. Alongside fault tickets recording remediation actions and return merchandise authorization (RMA) inspection reports, this heterogeneous operational dataset constitutes the data foundation of OptiMIND.

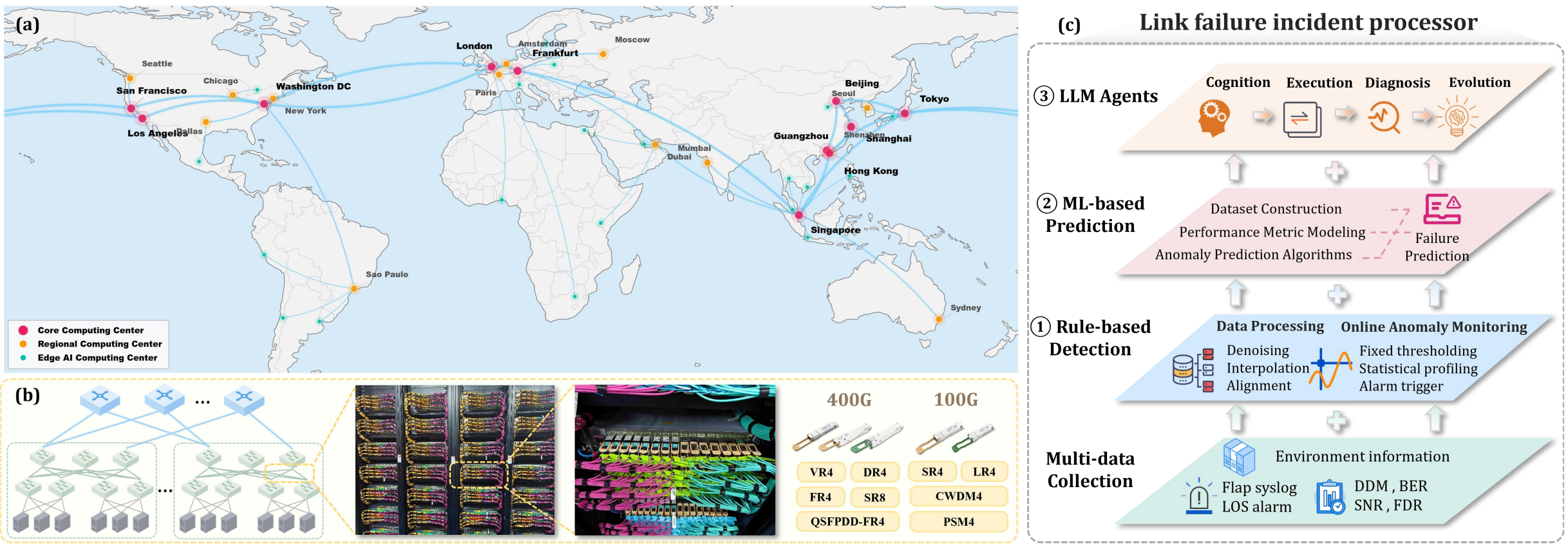

**Fig. 1:** Overview of the production AIDC network architecture. (a) Global DCI network. (b) Overview of DCN optical transceiver deployment. (c) The link failure incident processor.

We adopt a layered architecture that decouples deterministic anomaly screening from LLM-based reasoning. Specifically, as depicted in Fig. 1(c), a link failure incident processor narrows the search space from fleet-scale links to a small set of high-risk candidates through three stages: ① Rule-based Detection: raw metrics are filtered to identify samples exceeding predefined thresholds or exhibiting anomalous distributions; ② Machine Learning (ML)-based Fault Prediction: building upon our prior Future-Guided Learning model [7], we perform link-level prioritization through specialized prediction models; and ③ Description Aggregation for Agents: identified anomalies are packaged into structured fault cases with contextual metadata, upon which the multi-agent system performs causal reasoning and generates actionable diagnostic conclusions.

## The OptiMIND Framework

### *The Multi-Agent System*

OptiMIND employs a multi-agent architecture, as depicted in Fig. 2. The system adopts a Plan-and-Act paradigm [8] in which each agent is configured with a role-specific profile and accesses domain context via retrieval-augmented generation (RAG) from a persistent memory system incorporating Operations and Maintenance (O&M) knowledge, standard operating procedures (SOPs), and operational fault records. Specifically, the Planner acts as the central orchestrator. It receives risk link cases from the upstream data processor, decomposes them into dependent sub-tasks, and dispatches each to appropriate specialist agents. Three categories of sub-tasks are defined to cover the full fault management lifecycle: the Evaluator for fault risk assessment and prediction accuracy monitoring, the Executor for link failure localization and remediation suggestion, and the Diagnoser for failure mode classification, fault pattern correlation, and root cause diagnosis. Each agent is harnessed with a dedicated set of domain skills accessible through model context protocol (MCP) [9], and critical operations are subject to human-in-the-loop review. All structured outputs are forwarded to the Reflector, which performs multi-dimensional validation against production evidence and synthesizes verified results into consolidated diagnostic reports. Resolved cases are persisted within the memory system, progressively enriching the episodic knowledge base to serve as the foundation of the self-evolving mechanism.

### *The Self-Evolving Mechanism*

To continuously enhance the diagnostic capability of OptiMIND, we design a self-evolving mechanism, as shown in Fig. 2 (pink). This mechanism combines domain-adapted SFT for cold-start initialization with a continuously enriched memory system [10] for subsequent evolution. During the cold-start phase, we construct a high-quality SFT dataset of ~1,000 production fault cases, each annotated with chain-of-thought (CoT) rationales [11]. This dataset is used to fine-tune a Policy LLM that is shared across all agents through role-specific prompts, enabling it to acquire foundational diagnostic reasoning over historical fault patterns. To enable further evolution from live inference outcomes, we propose a memory system that automatically reviews and persists resolved cases, where each outcome is validated against production-verified metrics covering prediction accuracy, remediation correctness, and diagnostic completeness. Guided by this operationally grounded feedback, the Reflector autonomously updates agent skills across the full optical link management lifecycle, triggers iteration of existing fault prediction algorithms, and incrementally enriches the retrieval corpus for subsequent agent reasoning. This closed-loop process establishes OptiMIND as a robust, outcome-driven self-evolving system.

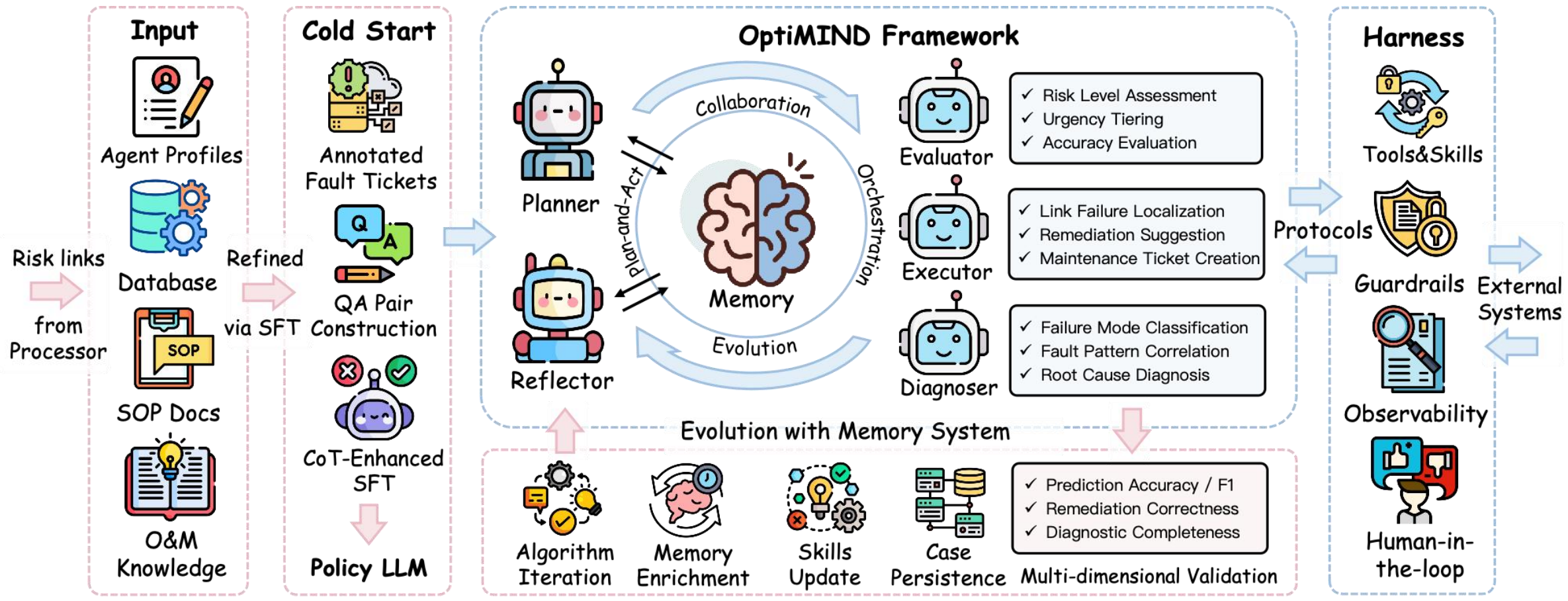

**Fig. 2:** Overview of the proposed OptiMIND framework: the multi-agent system (blue) and the self-evolving mechanism (pink).

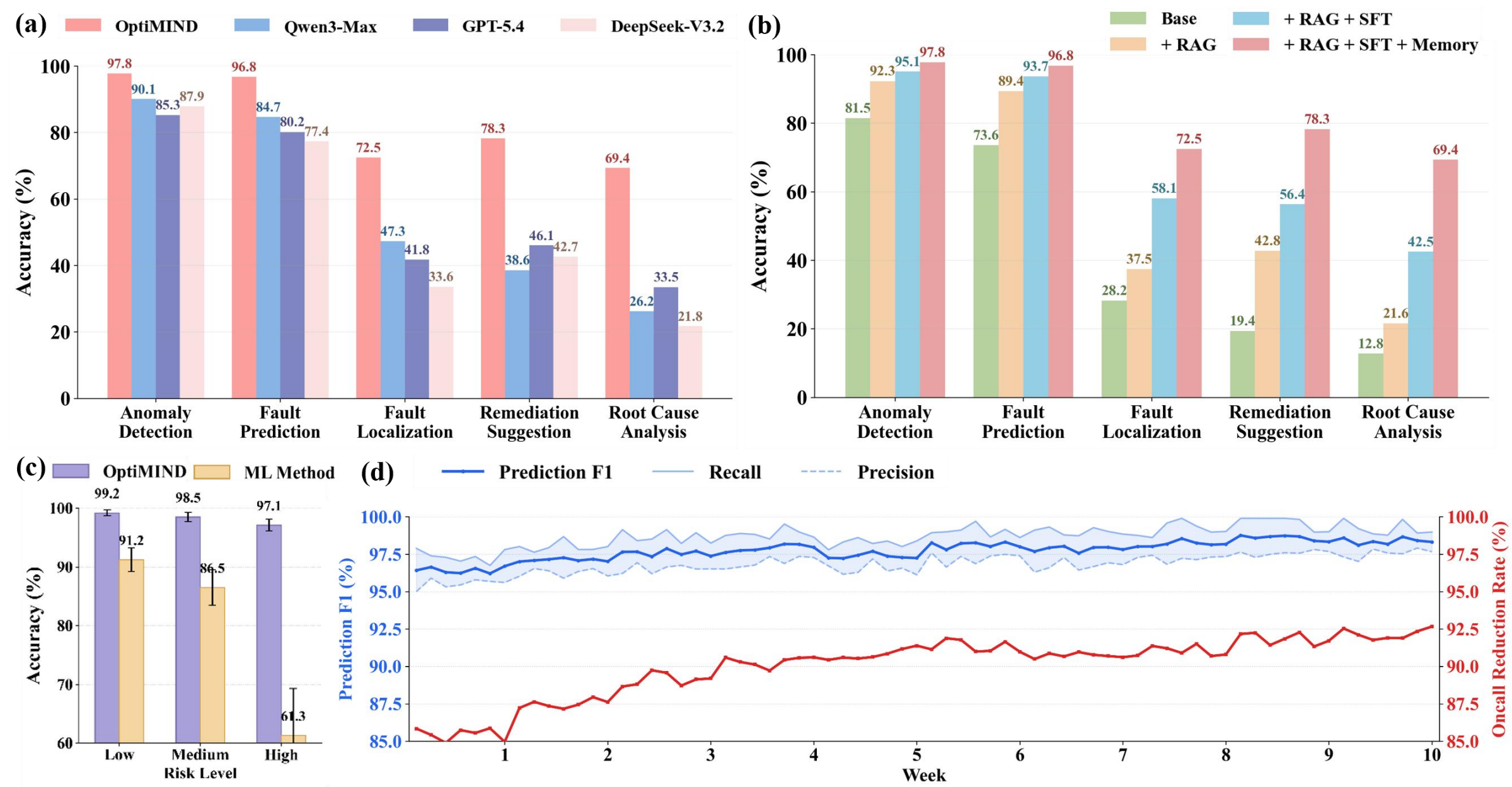

**Fig. 3:** Performance of OptiMIND. (a) Accuracy comparison with SOTA LLM baselines across core diagnostic tasks. (b) Ablation study on key components. (c) Fault prediction accuracy under different risk levels. (d) Self-evolving mechanism.

## Results

We collect seven months of multi-source telemetry data from Baidu's production AIDC infrastructure over millions of optical links. Each identified fault case in this dataset is paired with complete remediation tickets and root cause annotations. The dataset is used for SFT on Qwen3-8B [12] and initial knowledge base construction for the memory system. Fig. 3(a) illustrates the overall performance of OptiMIND compared with state-of-the-art (SOTA) general-purpose LLM baselines across five core optical link fault management tasks. OptiMIND achieves the highest accuracy on every task, with a particularly notable gain in root cause diagnosis, surpassing DeepSeek V3.2 by 47.6%. To validate the contribution of each core component, we conduct an ablation study. As shown in Fig. 3(b), the incorporation of RAG with a static knowledge base enhances the model's capability to capture contextual semantic associations and integrate external domain knowledge. Introducing domain-adapted paradigms via SFT equips the model with foundational reasoning capabilities, yet performance remains constrained. The proposed memory evolution mechanism further enhances adaptability to unfamiliar fault patterns by accumulating verified operational experience and recognized failure modes, thereby yielding significant improvements across all tasks.

To validate real-time inference performance, we evaluate OptiMIND on ten weeks of live network data. Fig. 3(c) compares its fault prediction accuracy against the previously deployed ML algorithms across different risk levels. OptiMIND exhibits significantly higher prediction and localization accuracy, particularly for high-risk optical link failures that have extensive impact. Furthermore, Fig. 3(d) illustrates the performance of OptiMIND throughout the ten-week period. It demonstrates a continuous self-evolving trajectory, with its prediction F1-score steadily reaching 97.7% and its failure incident reduction rate (FIRR) progressively ascending to over 60%—covering almost all port failures except for fan failures, power supply failures, and device crash errors.

## Conclusions

We present the first production network demonstration of an LLM-powered multi-agent system for autonomous management of millions of optical links in AIDCs. The system is refined through SFT on a seven-month operational dataset and continuously enriches its diagnostic memory through grounded feedback. Over a ten-week period of field validation across global DCNs, the system achieves a stable 97.7% prediction F1-score, providing a robust foundation for the reliable operation of large-scale AI infrastructure.

## Acknowledgements

This work was supported by Shanghai Pilot Program for Basic Research-Shanghai Jiao Tong University (21TQ1400213) and National Natural Science Foundation of China (62175145).